\documentclass[12pt]{amsart}
\usepackage{amsmath,amsthm,amssymb}
\usepackage[dvips]{graphics,epsfig}

\theoremstyle{definition}

\numberwithin{equation}{section}

\newcommand\cf{\textit{cf.}}
\newcommand\Z{{\mathbb Z}}

\newcommand\R{{\mathbb R}}

\newcommand\rank{\mathop{\text{rank}}}

\let\del=\partial

\newcommand\bkt[1]{\langle #1\rangle}
\newcommand{\spg}[2]{\{ #1 | #2 \}}	

\newcommand{\G}{\mathcal G}
\newcommand{\T}{\mathcal T}

\newcommand{\ivec}{\hat{\bold x}}
\newcommand{\jvec}{\hat{\bold y}}

\newcommand{\curl}{\mathbf{\nabla \times}}

\def\ttm#1{{\hbox{\ttfamily#1}}}	

\begin{document}
\renewcommand{\thefootnote}{\fnsymbol{footnote}}
\setcounter{footnote}{1}
\title[Cohomology for Anyone]{Cohomology for Anyone$^{\dagger}$}
\author{David A. Rabson}
\address{Physics Department, PHY 114 \\ University of South Florida \\ Tampa, FL 33617}
\email{davidra@ewald.cas.usf.edu}
\author{John F. Huesman}
\address{Physics Department, PHY 114 \\ University of South Florida \\ Tampa, FL 33617}
\email{jhuesman@ewald.cas.usf.edu}
\author{Benji N. Fisher}
\address{Mathematics Department \\ Boston College \\ Chestnut Hill, MA 02467}
\email{benji@member.AMS.org}
\urladdr{http://www2.bc.edu/$\sim$fisherbb}


\begin{abstract}
Crystallography has proven a rich source of ideas
over several centuries.  Among the many ways of looking at
space groups, N.\ David Mermin has pioneered the Fourier-space
approach.  Recently, we have supplemented this approach with
methods borrowed from algebraic topology.  We now show
what topology, which studies global properties of manifolds, has
to do with crystallography.
No mathematics is assumed beyond what the typical physics or
crystallography student will have seen of group theory; in particular,
the reader need not have any prior exposure to topology or to cohomology
of groups.

\end{abstract}

\maketitle

\footnotetext[2]{With apologies to N.D.\ Mermin, \cite{mermin81a,mermin81b}.}

\section{Introduction}
\renewcommand{\thefootnote}{\arabic{footnote}}
\setcounter{footnote}{0}

Reviewing the Fourier-space
formulation of crystallography, David Mermin wrote in 1992 that

\begin{quote}
        More than one person has told me that what I am calculating
        here are cohomology groups.  I have found this information
        less valuable than M.~Jourdain found the news that he was
        speaking prose, but am too ignorant to state with confidence
        that this is
	not a useful point of view. \cite{Mermin92a}
\end{quote}

Applying cohomology, we have proven certain theorems \cite{%
rabsonfisher02,FisherRabson01,fisherrabsoninprep}
in more
generality than they were previously known, including to
cases with 46-fold \cite{MRW46}
and other exotic rotational symmetries.  Our purpose here, using minimal
jargon and mathematical machinery, is to explain just what the
homological point of view is.  Since this point of view represents,
in some historical sense, a marriage between group theory and
topology, and since most readers will already be familiar with the
crystallographic implications of the former, we will display some
easily appreciated topological analogues and offer a complete
example space-group calculation, without shortcuts, employing some of the
new tools.

Most concisely, topology is concerned with numbers that remain
invariant under transformations; for example, the number of holes
in a doughnut does not change under rotations, continuous stretching,
or twisting.  In physics, electric charge does not change
under a gauge transformation, field theories routinely refer to
scalars as ``charges,''
and the Kosterlitz-Thouless treatment
of two-dimensional phase transitions has given rise to the description
of vorticity as ``topological charge'' \cite{thouless90}.  In the
traditional, direct-space, formulation of crystallography, the space
group of a structure does not depend on the choice of origin (``setting''),
despite the fact that not all point-group elements need pass through the
same point in a unit cell.  There is a long history of the application
of group cohomology to describe this invariance
\cite{AJ-I,AJ-II,Schw,JJ,Hiller}.

Similarly, Fourier-space crystallography, as first formulated by
Bienenstock and Ewald \cite{B-E} and developed by Mermin and
collaborators \cite{RWM,Rabson91,Mermin91,Mermin92a,Mermin92b,Dr-M,lifshitz96},
admits certain quantities invariant under a simpler analogue (to be
defined below) of the electromagnetic gauge transformation.  In the
special case of a periodic crystal, this gauge transformation
can be described by a translation for each point-group symmetry.
In the initial work on Fourier-space
crystallography, tracking down these invariants was incidental to the
main task of classifying space groups; most of the time, the invariant
of a non-symmorphic space group corresponds to a necessary extinction
in diffraction (something that clearly should not depend on the
arbitrary choice of a gauge).
Mermin first noticed that
the invariants might not always be so simple in the very article
alluding to the Bourgeois Gentleman.\footnote{%
Two of us were among those who suggested to David Mermin
that Fourier-space
crystallography could be described in the language of group cohomology;
Andr\'e LeClair was another
\cite{leclair_private_91}.  As far as we can tell, the existence
of this connection between Fourier-space crystallography and group
cohomology was first mentioned in print in Mermin's Moli\`ere reference;
in the same year, Piunikhin alluded briefly to the same thing
\cite{Piun-relation}.  While it is perhaps {\itshape interesting\/} that
results first formulated in one language can be recast in another,
the correspondence becomes {\itshape useful\/} when it leads to new results.}
There, he noted that of the 157 non-symmorphic periodic space groups in 
three dimensions, and additionally
of an important infinite class of quasiperiodic space groups that
includes all the quasicrystals so far discovered, only two (both among the
157) had
invariants that did {\itshape not\/} correspond to extinctions in
diffraction.\footnote{The absence of systematic extinctions in these
two space groups, $I2_12_12_1$ and $I2_13$, was well known.  What
was new was the description of an algebraic invariant that ``detects'' that
these two groups are non-symmorphic.}
Later, Mermin and K\"onig \cite{K-M:1997,K-M:1999,K-M:2000} showed
that the new kind of invariant had a different physical interpretation,
that of electronic ``band sticking.''  The K\"onig-Mermin conditions
on the invariant for band sticking were too restrictive; we have
generalized them \cite{FisherRabson01} and have also found a third
type of invariant corresponding to neither extinctions nor
K\"onig-Mermin band sticking \cite{fisherrabsoninprep}.

\subsection{Fourier-Space Crystallography}
\label{section:FSC}
In order to establish notation, we briefly review the
Rokhsar-Wright-Mermin formulation of crystallography in Fourier space; see
reference \cite{Dr-M} or \cite{lifshitz96} for a more careful development.
We begin with the (reciprocal)
lattice $L$, which consists of all integral linear combinations
of a finite number of generating vectors.\footnote{%
The lattice of a periodic structure is discrete and generated
by three vectors in three dimensions, while that of a quasiperiodic structure
is not discrete and has more generators than dimensions.  In Fourier-space
crystallography, this is the only theoretical distinction between
these two cases.  Since we are not interested in the direct-space lattice
(if it even exists), we shall use the term ``lattice'' for the
the reciprocal lattice.}
The Fourier transform of a density (mass, electronic, {\itshape etc.}) has
support on this lattice; we will refer to the Fourier transform as the
density, since the direct-space density plays no role in the theory.
Two densities, $\rho(\bold k)$ and $\rho'(\bold k)$,
are indistinguishable if all their
$n$-body correlation functions are the same; this is equivalent to the
condition
\begin{equation}
\rho'(\bold k)=e^{2\pi i\chi(\bold k)}\rho(\bold k)\rlap{\quad,}
\label{densities}
\end{equation}
where the {\itshape gauge function\/} $\chi(\bold k)$, defined modulo unity,
is linear on the lattice.  In case a point operation $g$ takes $\rho$
to $\rho'$, we write the special gauge function associated with $g$
as the {\itshape phase function} $\Phi_g$; the point group $G$ consists of
all $g$ such that
\begin{equation}
\rho(g\bold k)=e^{2\pi i\Phi_g(\bold k)}\rho(\bold k)\rlap{\quad.}
\label{phase}
\end{equation}
The phase function is the central object of the Fourier formulation, and
it is subject to a restriction and to an identification.
As a consequence of the associativity of the group action on the lattice,
the phase function must satisfy the {\itshape group-compatibility condition\/}
for $g,h\in G$,
\begin{equation}
\Phi_{gh}(\bold k)=\Phi_g(h\bold k)+\Phi_h(\bold k)\rlap{\quad,}
\label{compat}
\end{equation}
while two phase functions, $\Phi$ and $\Phi'$, describe indistinguishable
densities if they are related by a gauge function, $\chi$,
\begin{equation}
\Phi'_g(\bold k) = \Phi_g(\bold k) +
\chi(g\bold k-\bold k)\rlap{\quad,}
\label{gauge}
\end{equation}
and so are identified.
The lattice~$L$, the action of the point group~$G$, and the phase
function~$\Phi$ determine the \textit{space~group type} of the crystal.
Homology theory provides
a convenient way of calculating all such functions, subject to the
restriction \eqref{compat} and the identification \eqref{gauge}.

\section{Invariants}
We continue the review of Fourier-space crystallography, stressing
the role played by gauge-invariant values of combinations of phase
functions.  We then switch context to elementary topology, where
we define homology of loops drawn on two-dimensional surfaces
in an intuitive way.  Cohomology is introduced through
a familiar example from vector calculus. Finally, we connect
topology to crystallography by way of homology and cohomology of groups.

\subsection{Invariants in Fourier-space Crystallography}
\label{invariantsFS}
Evidently, gauge-equivalent phase functions $\Phi$ and $\Phi'$
in \eqref{gauge}
will have different values when evaluated for generic $g$ and $\bold k$.
However,
certain linear combinations of phase functions are independent of
gauge.  Immediately from \eqref{gauge}, we see that if $g$ leaves
$\bold k$ invariant, then $\Phi_g(\bold k)$ is the same in any gauge;
we call such a quantity a {\itshape gauge invariant of the first kind}.
It follows from \eqref{phase} that if a gauge invariant
of the first kind is non-zero (always modulo the integers), then
$\rho(\bold k)=0$, so there is an extinction in diffraction.
For example, let
$ G = 2mm = \{e, r, m, rm\} $,
where $e$ is the identity,
$r$~denotes a $180^\circ$~rotation, and $m$~is reflection in the
$\hat{\bold x}$-axis;
let $L$~be the lattice generated by vectors $a$ and $b$ along
the $\hat{\bold x}$- and $\hat{\bold y}$-axes.
Then it is not hard to check that
$ \Phi_m( i a + j b) = 
  \Phi_{rm}( i a + j b) = i \tfrac12
$,
$ \Phi_e(\bold k) = \Phi_r(\bold k) = 0 $
satisfies the group-compatibility condition~\eqref{compat}.
Thus $\Phi_m(a)=\tfrac12$ is an invariant of the first kind, and
the point $a$ is extinct in diffraction.\footnote{%
The point $a$ may also be extinct for the point group $4mm$
on the square lattice (and for analogous star lattices $8,\,16,\dots$).
That there is {\itshape no\/}
systematic extinction for $6mm$ on the triangular lattice, as a consequence
of \eqref{compat}, demonstrates that the calculation of 
space groups is non-trivial.}
For further discussion of this example, see \S\ref{connection}
below.

This is not the only kind of invariant.  Of course any integral
linear combination (such as $\Phi_{g_1}(\bold k_1) + \Phi_{g_2}(\bold k_2)$
where $g_i\bold k_i=\bold k_i$) of gauge invariants of the first kind is still
an invariant, but for two of the 157 periodic non-symmorphic
space groups in three dimensions, the simplest gauge-invariant
quantity one can construct takes the form
\begin{equation}
\Phi_g(\bold k_h) - \Phi_h(\bold k_g)\rlap{\quad,}
\label{2KM}
\end{equation}
where $g$ and $h$ commute and
where neither term
alone is gauge invariant.  K\"onig and Mermin \cite{K-M:1997,K-M:1999,K-M:2000}
define the lattice vectors $k_g$ and $k_h$ in terms of a point $\bold q$
{\itshape not\/} in the reciprocal lattice but with the property that
\begin{equation}
\bold k_g~\equiv~\bold q-g\bold q
\quad \text{and} \quad
\bold k_h~\equiv~\bold q-h\bold q
\label{ka}
\end{equation}
{\itshape are\/}.  The group operations $g$ and $h$ are then elements of
the {\itshape little group\/} of $\bold q$.  We refer the reader to their
papers or to \cite{rabsonfisher02} for the proof that if the invariant
\eqref{ka} is non-zero, {\itshape any\/} electronic energy level at wavevector
$\bold q$ must be at least two-fold degenerate.\footnote{Equation
\eqref{2KM} is a special case of what we refer to in
\cite{FisherRabson01,fisherrabsoninprep} as an
invariant of the second kind, which always leads to
the electronic degeneracy;
there also exists an invariant of the
third kind, but we shall not need it here.}

\addtocounter{footnote}{1}
\subsection{Topological Invariants$^{\thefootnote}$}\relax
\footnotetext{The elementary treatment of topology in this section
draws on Alexandroff's slim introduction
\cite{alexandroff61}.}
\label{topo}
Figure 1 offers a tour of some of the objects of topology.
The shaded area represents a two-dimensional manifold $M$.  The region
$S$ is a submanifold; the other letters label various oriented curves.
An oriented curve is called a 1-simplex.
We have an intuitive idea, which we shall make formal shortly, of
what it means for objects to bound one another.  As examples, the $1$-simplex
$AB$ is bounded by the points $A$ and $B$, but the other labeled $1$-simplices,
which
are closed, have no boundary points; such 1-simplices with no boundaries will
be examples of $1$-cycles.
Cycles themselves may or may not bound submanifolds:
the $1$-cycle $C$ bounds
$S$, but $D$ is not the boundary of any submanifold of $M$.
To describe the boundary of $M$, we must exclude the two holes
and so need a combination of cycles, which we shall see can be written
as the sum $H-F-G$.  Such a formal sum of $1$-simplices is called a $1$-chain;
any sum, such as $C+C+AB=2C+AB$, may be performed,
although the result might not be very interesting.

\def\captionone{%
The manifold $M$, containing two holes, illustrates a
topological space.  $AB$, $C$, $D$, $E$, $F$, $G$, and $H$ are
oriented $1$-simplices; of these, all but $AB$ are $1$-cycles.
The $1$-cycles $C$ and $E$ are $1$-boundaries.  In particular,
$C$ is the boundary of sub-manifold $S$.
}
\begin{figure}[tp]
\centerline{\epsfxsize0.65\hsize\epsfbox{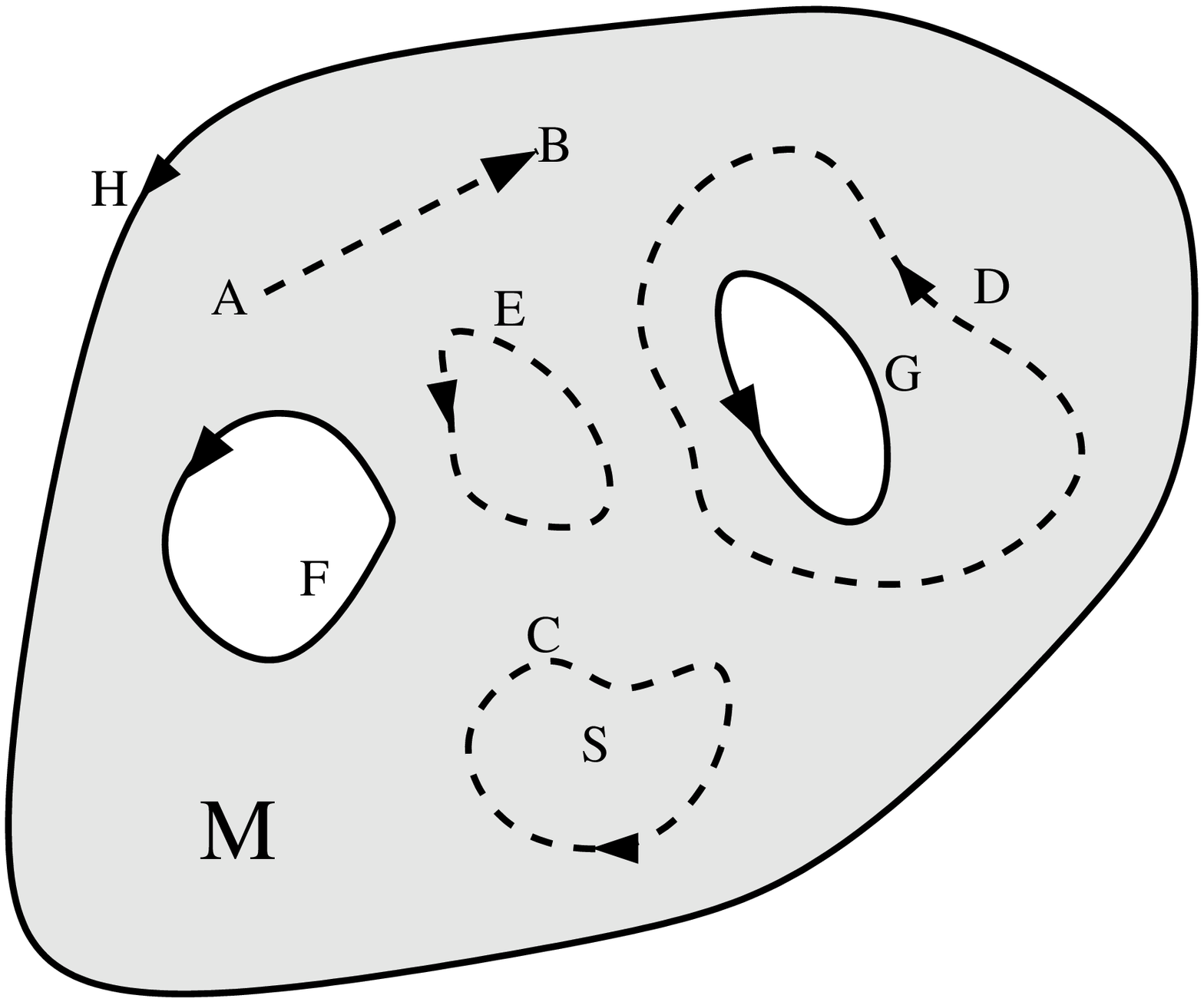}}
\caption[Figure 1]{\captionone}
\end{figure}

The cycle $F$ encloses the left hole once, while cycles like $D$ and
$G$ each enclose the right hole once.  A cycle might
enclose one or both holes an arbitrary number of times; we could draw
a lemniscate (figure-eight) enclosing the left hole $-1$ times and
the right hole $+1$~times.

Boundaries offer a way to identify cycles, such as $D$ and $G$, that
enclose holes the same number of times.  Since $D-G$
bounds the region between the two cycles, we shall write $D=G+(D-G)~\doteq~G$,
identifying the two cycles because they differ by the boundary $D-G$.
Two cycles differing by
a boundary will be called \textit{homologous}; cycles $E$ and $C$ are also
homologous to each other but not to $D$ and $G$.
A set of homologous cycles forms a \textit{homology class}; the enumeration of
classes of cycles identified by homology will have a direct analogue
in the classification of phase functions identified by gauge transformation.

For this example, each homology
class is labeled by two integers:  one describing how many times each
cycle in the class goes around the left hole and another
describing how many times each cycle in the class goes around the
right.  Since cycles form an additive group, so do homology classes;
the homology group of the figure\footnote{The
homology group of these curves is called the first homology group; elsewhere
\cite{FisherRabson01,rabsonfisher02} we have had occasion to employ
higher homology groups.  We use the common mathematical notations
$\Z$ for the integers, $\R$ for the real numbers.} is $\Z\times\Z$.

The torus, $T$, of Figure~2a (the surface of a doughnut, not the dough) cannot
be deformed into the manifold of Figure~1, yet the two have the same
homology group. This will follow from the fact that a cycle on the torus
may enclose the hole any integer number of times and that it may enclose
the dough (which is also a kind of hole) any integer number of times.
Since this assertion is perhaps a little less transparent than the
corresponding one for Figure~1, we
start by describing a general method for calculating homology groups.

\def\captiontwo{%
Figure 2a illustrates the torus $T$ (doughnut) embedded in Euclidean
3-space.  Only the surface forms $T$.  Figure 2b illustrates the same
torus: the sides with double arrows are to be joined, generating a
cylinder, and then the sides with single arrows.  The triangulation
consists of two $2$-simplices, $\Delta$ and $\Gamma$, three $1$-simplices,
$b$, $c$, and $d$, and just a single $0$-simplex (vertex), $A$.  The
$1$-cycles $b$ and $c$ are also shown in Figure 2a.
}
\begin{figure}[tp]
\newbox\foobox\setbox\foobox\vbox{\hsize.35\hsize%
\epsfxsize\hsize\epsfbox{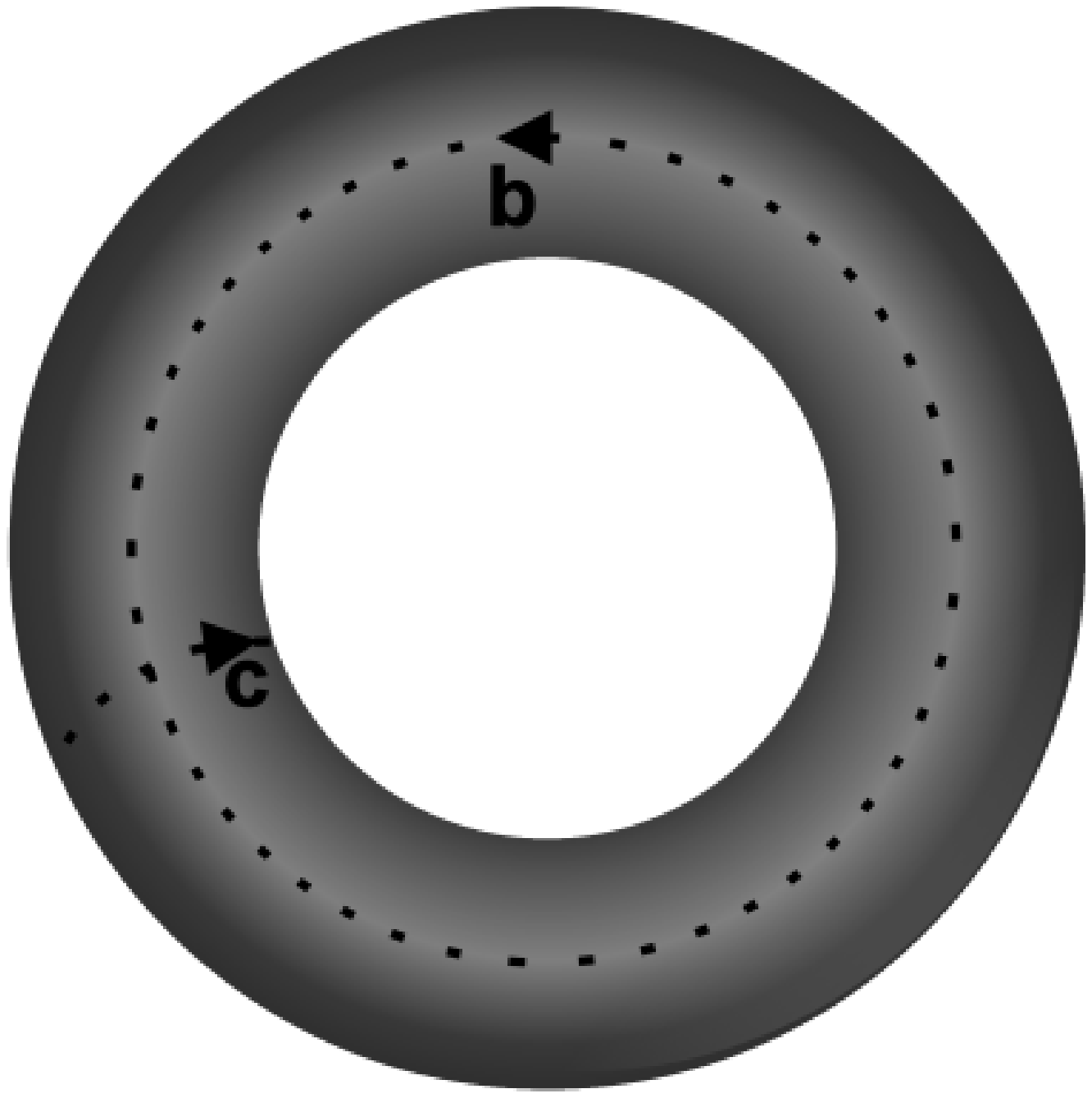}}
\newbox\barbox\setbox\barbox\vbox to\ht\foobox{\hsize.4\hsize
\epsfxsize\hsize\hbox{\raise.9\baselineskip\hbox{\epsfbox{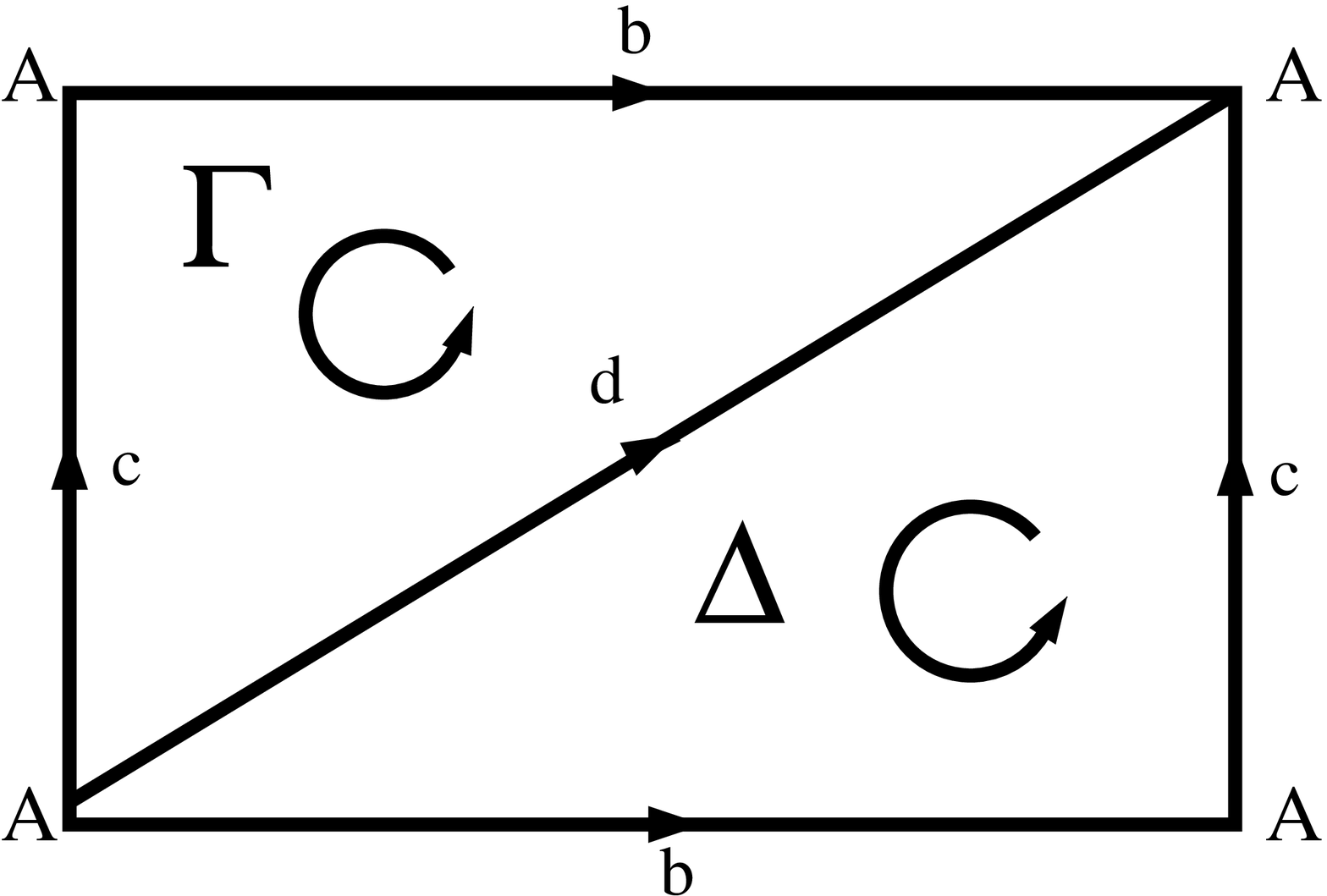}}}\vfil}
\centerline{\hfil%
\vbox{\unvcopy\foobox\vskip0.5\baselineskip\hbox to\wd\foobox{\hfil(a)\hfil}}%
\hfil\hfil
\vbox{\unvcopy\barbox%
\vskip0.5\baselineskip\hbox to\wd\barbox{\hfil(b)\hfil}}%
\hfil
}
\medskip
\caption[Figure 2]{\captiontwo}
\end{figure}

First we flatten the torus into the
rectangle of Figure~2b.
The arrows indicate
that opposite sides are to be identified, or sealed together: after
identifying the two sides with double arrows, one obtains a cylinder;
identifying the other two sides then gives the torus.  Figure 2b shows the
torus
with a diagonal added.  In this
triangulation, the torus $T$ consists of a single $0$-simplex, or point,
labeled $A$; the three $1$-simplices $b$, $c$, and $d$; and the two
$2$-simplices (triangles) $\Delta$ and $\Gamma$.  The straight arrows on the
$1$-simplices indicate their orientations; $2$-simplices can be oriented
as well, as indicated by the curly arrows.  The orientations of the
$2$-simplices and their constituent $1$-simplex sides
bear no necessary relation.
Neither the triangulation of $T$ nor the
assignment of orientations
is unique; we refer the reader to \cite{alexandroff61}
for theorems guaranteeing that the homology group is independent of
these choices.

Labeling a $1$-simplex by its endpoints (which may be the same point)
and a $2$-simplex (triangle) by its vertices,%
\footnote{This labeling is not meant to suggest that a simplex is determined 
by its vertices.  The \emph{order} of the vertices does determine the
orientation of the simplex.}
we define the \textit{boundary operator} $\partial$ by
\begin{equation}
\begin{aligned}
\partial(x_0,x_1) &= x_1 - x_0\\
\mbox{and\quad}\partial(x_0,x_1,x_2) &= 
(x_0,x_1) - (x_0,x_2) + (x_1,x_2)\rlap{\quad.}
\end{aligned}
\label{partial1}
\end{equation}
For example, in Figure~2a,
$ \del b = \del c = \del d = 0 $,
and
\begin{equation}
\begin{aligned}
\partial\Gamma &= -b -c + d \\
\partial\Delta &= b + c -d\rlap{\quad.}
\end{aligned}
\relax\label{GD}
\end{equation}

The definitions extend to $1$-chains and $2$-chains by linearity,
and they agree with the intuitive notion of boundary.  So the
$1$-simplex~$(x_0, x_1)$ is a $1$-cycle---that is, $\del (x_0, x_1) = 0$---if
and only if
$ x_0 = x_1 $.
The boundary of a triangle (Figure 3)
$(x_0,\, x_1,\, x_2)$  is the sum of its sides, if we accept the convention
that $(x_0,\,x_2)\doteq-(x_2,x_0)$.

\def\captionthree{%
The boundary of a $2$-simplex (triangle) is the sum of its sides.
According to \eqref{partial1}, we have $(x_0,x_1) - (x_0,x_2) + (x_1,x_2)
=-(x_1,x_0)+(x_1,x_2)+(x_2,x_0)$, in agreement with the orientations
in the figure.
}
\begin{figure}[tp]
\centerline{\epsfxsize0.45\hsize\epsfbox{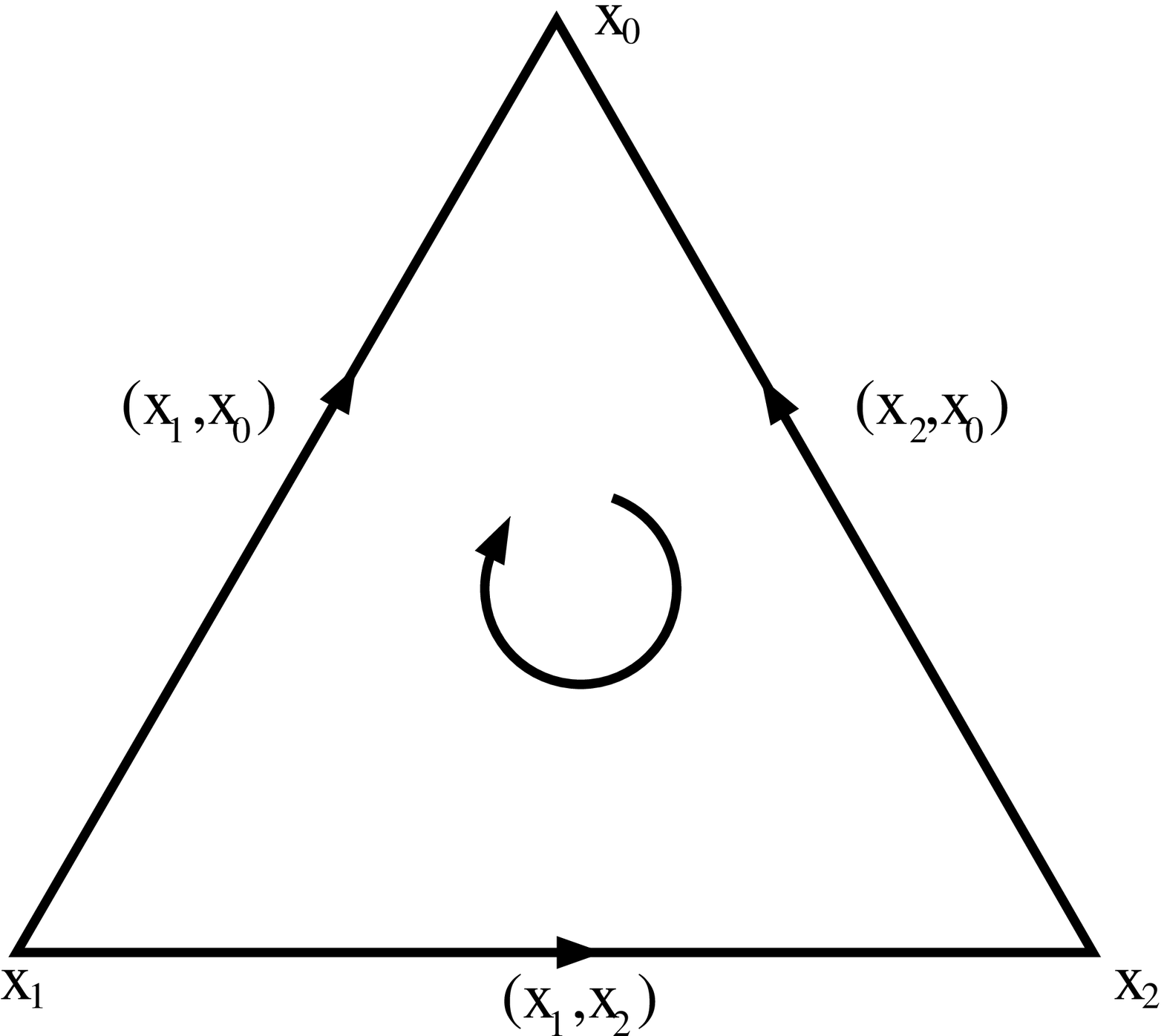}}
\caption[Figure 3]{\captionthree}
\end{figure}

It follows from these definitions that the boundary of the boundary of
a $2$-chain vanishes.  Since we may define the boundary of
a point (a $0$-simplex) as $0$, the boundary of the boundary of
a $1$-chain also vanishes. In fact, in the general theory,
$\partial\partial$ is identically zero.

The $1$-chains whose boundaries are $0$ form the group $Z_1$ of
\textit{$1$-cycles}, while the $1$-cycles that can themselves be written
as boundaries of $2$-chains constitute the group $B_1$ of
\textit{$1$-boundaries}.  Homology identifies $1$-cycles that
differ only by $1$-boundaries:  that is, we form the quotient
\begin{equation}
H_1 ~=~ Z_1 / B_1\rlap{\quad,}
\label{H1}
\end{equation}
called the first homology group.

Applying these ideas to the triangulation of the torus, we first
calculate the group $C_1$ of all $1$-chains.  Since there are exactly three
$1$-simplices, an arbitrary $1$-chain takes the form $\ell b + m c
+ n d$, with $\ell, m, n \in \Z$; thus $C_1$ is a three-dimensional
``space'' (properly, module) over the integers.
In this example,
all three $1$-simplices have vanishing boundaries,
so the group of $1$-cycles,
$Z_1$, is the same as $C_1$:
\begin{equation}
Z_1 = \langle b,\,c,\,d \rangle\rlap{\quad,}
\label{Z1gen}
\end{equation}
where the angle brackets indicate the generators (over the integers).
\relax From \eqref{GD},
we see that the only $1$-boundaries are integral multiples of
$b + c - d$, so that $B_1=\langle b+c-d\rangle$, a one-dimensional module.
As a group, first homology with coefficients in the integers is given by
\begin{equation}
H_1(T,\,\Z) ~=~ \Z^3/\Z ~=~ \Z\times\Z\rlap{\quad.}
\label{H1T}
\end{equation}
That is, any $1$-cycle is homologous to a linear combination of~$b$ (going
once around the hole) and~$c$ (going once around the dough),
as we claimed above.

We have taken the coefficients of simplices to be integers, but we
can use any additive group.  One natural choice is the integers
modulo two, $\Z_2=\{0,\,1\}$ with the addition rules $0+0=1+1=0$,\hbox{\ }
$0+1=1+0=1$.  Over $\Z_2$, twice any cycle vanishes.
This corresponds to unoriented simplices: for
example, if traversing the $1$-simplex $C$ in Figure~1 once
is identified with traversing the
oppositely-oriented simplex once, we have $-C\doteq-C+2C=C$, so the
coefficients are in $\Z_2$.  If we use unoriented simplices,
the homology groups of $M$ and $T$ are both two copies of the
integers modulo 2:
\begin{equation}
H_1(M,\,\Z_2) ~=~ H_1(T,\,\Z_2) ~=~ \Z_2\times\Z_2\rlap{\quad.}
\label{H1MZ2H1TZ2}
\end{equation}

Before leaving the subject of manifolds, we consider the projective
plane, $\R P^2$, which can be represented, like the torus,
as a rectangle, but now opposite sides are identified in opposite directions,
as in Figure~4.%
\footnote{
Equivalently, it may be thought of as a disk with opposite points of
the circumference identified or, \textit{via} stereographic projection, as a
Euclidean plane plus a line at infinity.  Since each line contains a point at
infinity, \textit{projection} from the focal point~$P$ to a line not passing
through~$P$ is defined on the entire plane except for~$P$, hence the term
\textit{projective plane}.}
The triangulation of the figure comprises
the two $0$-simplices $P$ and $Q$, three $1$-simplices, $f$, $g$, and $h$,
and two $2$-simplices, $\Upsilon$ and $\Psi$.  It is easy to see that
all cycles are generated by $h$ and $f+g$.
We determine the boundaries by
\begin{equation}
\begin{aligned}
\partial\Upsilon &= h -f -g \\
\partial\Psi &= -h -f -g\rlap{\quad.}
\end{aligned}
\label{UpsilonPsi}
\end{equation}
Thus $2h=\partial\Upsilon-\partial\Psi$
is a boundary, but the cycle $h$ is not.  The other generating
cycle, $f+g$, differs from $h$ by the boundary $\partial\Upsilon$;
so there exists only a single cycle, which we may take to be $h$,
modulo boundaries.
The first homology group is
\begin{equation}
H_1(\R P^2,\,\Z)~=~H_1(\R P^2,\,\Z_2)~=~\Z_2\rlap{\quad.}
\label{H1RP2}
\end{equation}

\def\captionfour{%
The projective plane $\R P^2$ (compare Figure 2b).  Now, opposite sides are to
be twisted before being glued together; we despair of showing what the
result might look like in four dimensions.  This  triangulation consists
of two $2$-simplices, $\Psi$ and $\Upsilon$, three $1$-simplices, $f$,
$g$, and $h$, and two $0$-simplices, $P$ and $Q$.  Twice $h$ is a boundary,
and modulo boundaries, $h$ is the only cycle.
}
\begin{figure}[tp]
\centerline{\epsfxsize0.5\hsize\epsfbox{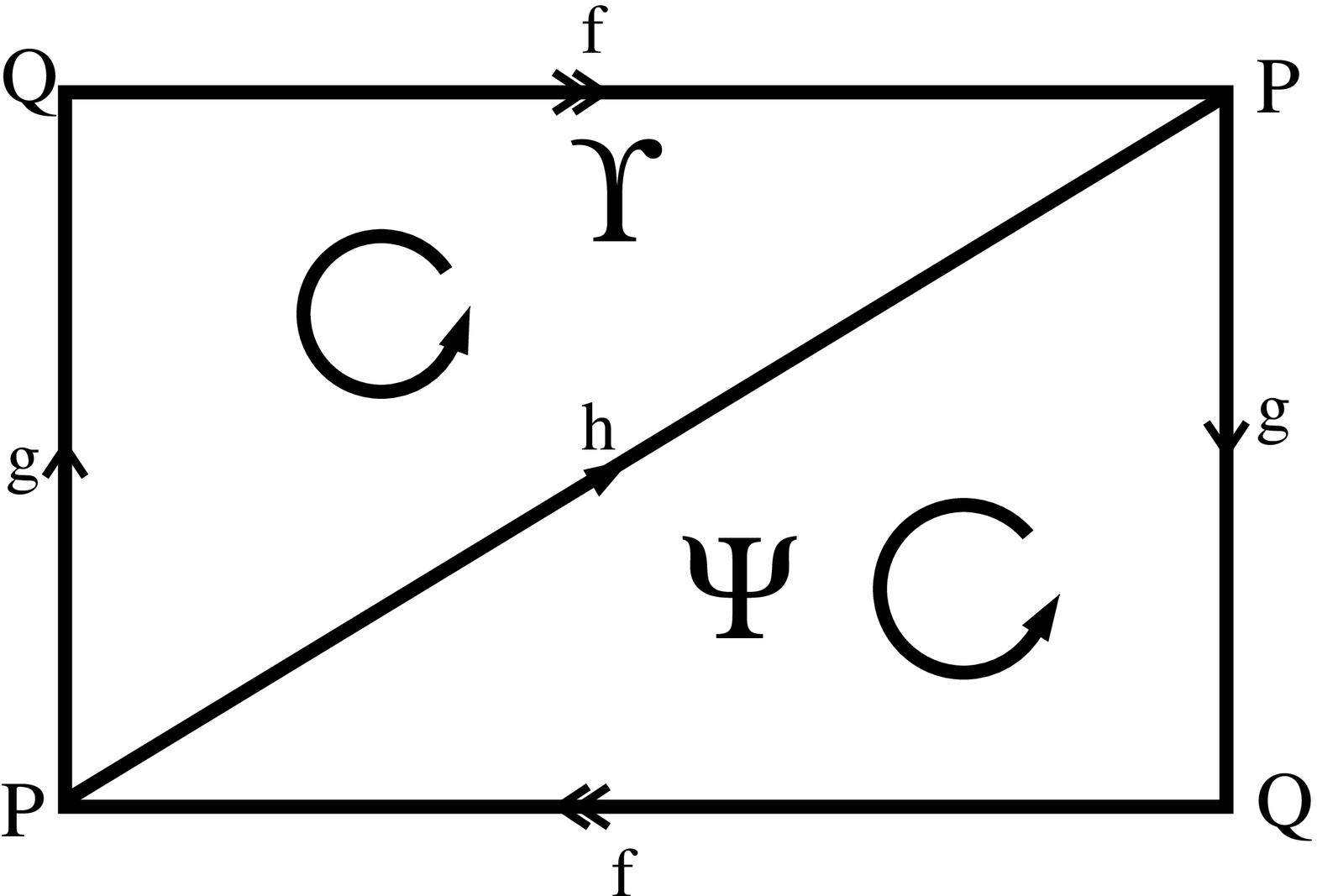}}
\caption[Figure 4]{\captionfour}
\end{figure}

The fact that any triangulation can be used to calculate the homology of a
manifold is related to Euler's formula:
\begin{equation}
	\label{eqn:Euler}
	V - E + F = 2 ,
\end{equation}
where $V$,~$E$, and~$F$ denote the number of vertices, edges, and faces (or
0-, 1-, and 2-simplices) in a polyhedron.  A polyhedron can be thought of as a
polygonal dissection of the sphere; if we want, we can triangulate the faces
and so obtain a triangulation of the sphere.  The triangulation (or the
polygonal dissection, if one takes the effort to define boundaries of faces
with more than three sides) can be used to compute the homology groups 
of the
sphere~$S$, and in particular the alternating sum
\begin{equation}
	\label{eqn:EulerChar}
	\chi(S) = \rank H_0(S) - \rank H_1(S) + \rank H_2(S) ,
\end{equation}
known as the \textit{Euler Characteristic} of the sphere.  It follows from
elementary linear algebra that the alternating sum is the same whether one
takes ranks of homology groups or of chain groups:  that is,
$ \chi(S) = V - E + F $
is the same for any dissection of the sphere.  From the triangulations above, 
it follows that the
Euler characteristics of the torus and projective plane are $0$ and $1$.

\subsection{From cycles to vector fields and de Rham cohomology}
\label{deRham}

Homology of a manifold is closely connected to the question of which
vector fields are conservative.  Let
$ \mathbf{F} = P\ivec + Q\jvec $
be a vector field in the plane.
If $\mathbf{F}$ is the gradient of a potential~$\phi$, then
the rules of vector calculus imply that $\curl\mathbf{F}=\mathbf{0}$
and that the line integral of $\mathbf{F}$ over any contour vanishes.
On the other hand, a curlless
vector field
may or may not be the gradient of a potential.

The standard counterexample is
\begin{equation}
	\label{eqn:dtheta}
	\mathbf{F} = \frac{x \jvec -y \ivec }{x^2 + y^2} .
\end{equation}
It is easy to verify that
$ \curl \mathbf{F} = \mathbf{0} $
but that integrating counterclockwise around
the unit circle~$C$ gives
$ \oint_C \mathbf{F} \cdot d \mathbf{r} = 2 \pi $.
We could say that $\mathbf{F}$ has a $\delta$-function curl at the origin
or, equivalently, we could cut the origin out of the plane, yielding
a ``punctured plane,'' $M$.
In
the language of homology, the circle~$C$ is a cycle in $M$
but not a boundary.  If, instead of~$C$, we consider the boundary~$B$ of a
region~$R$ that does not contain the origin, then Green's Theorem implies that
$\oint_B\mathbf{F}\cdot d\mathbf{r} = 0$.

More generally, consider a vector field that is curlless on
a plane punctured any number of times.  Two cycles
that differ by a boundary yield the same line integral, so homology
naturally describes the classes of line integrals of a curlless vector field.

It is equally natural to fix a cycle and to consider different
vector fields.  Adding the gradient of a potential to $\mathbf{F}$ does
not change any closed line integral, so we identify two vector fields
that differ by a conservative field.  
The resulting vector space is the first \textit{de~Rham cohomology group} of
the punctured plane, $M$, denoted
$ H^1_{\text{DR}}(M) $.

The condition (that $\mathbf{F}$ have no curl on $M$) and the identification
(of vector fields differing by conservative fields) mirror the cycle and
boundary conditions on contours, providing
a natural duality between de~Rham cohomology and
$ H_1(M,\R) $,
the homology of~$M$ with real (not integral) coefficients.  On the level of
chains and vector fields, the duality is defined by the circulation integral,
$ \bkt{c,\mathbf{F}} = \oint_c \mathbf{F} \cdot d \mathbf{r} $.
To see that this is well defined on (co)homology, it is enough to
notice that
$ \oint_c \mathbf{F} \cdot d \mathbf{r} = 0 $
if $c$~is any cycle and $\mathbf{F}$~is conservative (fundamental theorem of
calculus) or if $c$~is a boundary and $\mathbf{F}$~is irrotational (Green's
Theorem).

The language of differential forms \cite{spivak} connects the gradient
and the curl.
In this language,
$\omega\equiv\mathbf{F}\cdot d\mathbf{r}=P\,dx+Q\,dy$
is a differential $1$-form.
The differential of $\omega$, giving the curl,
$d\omega=(\partial Q/\partial x
-\partial Q/\partial y)\,dx\,dy$, is a $2$-form.  The exterior derivative~$d$
serves as the {\itshape coboundary} operator in this context;
the vanishing of
$d\omega$ makes $\omega$ a {\itshape $1$-cocycle}.
This resembles the cycle condition in homology, except that
where $\partial$ demoted an $n$-chain to an $n-1$-chain, here $d$
promotes a $1$-form into a $2$-form.
Similarly,
a $0$-form is a scalar-valued function, $\phi$, on the plane, and
$d\phi=(\partial\phi/\partial x) dx + (\partial\phi/\partial y) dy$
is a {\itshape $1$-coboundary}.  De Rham cohomology is the space of
cocycles modulo coboundaries, {\itshape i.e.}, vector fields curlless on
$M$ modulo conservative fields.\footnote{We note that this
same example has an equivalent interpretation with
contour integrals over functions that are analytic in the complex plane
except at simple poles.
}
Table 1 summarizes some of the relations between homology and de Rham
cohomology.
\begin{table}
\begin{center}
\renewcommand{\arraystretch}{1.1}
\tabcolsep1.5\tabcolsep
\begin{tabular}{|c|c|c|}
\hline
& homology & cohomology \\
\hline
$1$-(co)chain & sum of curves on $M$ & vector field, $\bold F$ \\
$1$-(co)cycle & closed curves (contours for $\oint$) & $\nabla\times\bold F=0~$ on $M$ \\
$1$-(co)boundary & trivial: bounds region of $M$ & trivial: $\oint\bold F\cdot d\bold r=0$ \\
\hline
\end{tabular}
\end{center}
\caption[Table 1]{Summary of some of the relations between homology
and de Rham cohomology.}
\end{table}

\subsection{Crystallographic Invariants as Homology Elements}
\label{section:duality}
In the foregoing, we identified closed curves (cycles)
differing only by boundaries.
We then enumerated all the
possible cycles, up to boundaries.
We now describe a similar way to find all crystallographic invariants,
demonstrating the method in the next section.

First, instead of thinking of a collection of functions~$\Phi_g$, one for each
element of the point group, think of the phase function~$\Phi$ as acting on a
pair consisting of a point-group element $g$ and a lattice vector $\bold k$.
We will write this pair as $\bold k[g]$; such an object is an example of a
\textit{$1$-chain}.  The phase function acts on this $1$-chain in the obvious way:
$\Phi(\bold k[g])=\Phi_g(\bold k)$.  Addition is defined on
$1$-chains as though the various elements of the point group $G$ were
independent basis vectors in a vector space and with the reciprocal
lattice playing the role of numeric coefficients.\footnote{Since the lattice
is not a field, the $1$-chains form a module instead of a vector space.}
Since the phase function is linear on the lattice, it acts
distributively over addition.  Thus, for example,
\begin{equation}
\Phi(\bold a[g] + \bold b[g] + \bold c[h])
~=~ \Phi(({\bold a+\bold b})[g] + \bold c[h])
~=~ \Phi_g(\bold a+\bold b) + \Phi_h(\bold c)\rlap{\quad.}
\end{equation}

Define the boundary operator on a $1$-chain by analogy to \eqref{partial1}:
\begin{equation}
\partial\bold k[g] = g\bold k - \bold k
\label{partialkg}
\end{equation}
so that for a general $1$-chain $c=\sum_i\bold k_i[g_i]$,
\begin{equation}
\partial c = \sum_i g_i\bold k_i - \bold k_i\rlap{\quad.}
\label{delc}
\end{equation}

A $1$-chain $c$ is a $1$-cycle if $\partial c=0$.  Considering the
gauge transformation of \eqref{gauge}, we have for the difference between
phase functions $\Phi$ and $\Phi'$ evaluated on the $1$-cycle $c$
\begin{equation}
\Phi'(c)-\Phi(c) ~=~ \chi\Bigl(\sum_i g\bold k_i-\bold k_i\Bigr)
~=~ 0\rlap{\quad.}
\label{gaugeinvariant}
\end{equation}
Thus cycles are gauge invariants, and since $\chi$ is an arbitrary
linear function (from the lattice to the real numbers modulo unity),
{\itshape any\/} gauge invariant is a cycle.

Certain gauge invariants are trivial.  Applying \eqref{partialkg}, we
establish that any $1$-chain
of the form
\begin{equation}
b ~=~ h\bold k[g] - \bold k[gh] + \bold k[h]
\label{trivialcycle}
\end{equation}
is a $1$-cycle.  However, {\itshape any\/} phase function
$\Phi$ evaluated at $b$ must vanish by the group-compatibility
condition \eqref{compat}.  Adding $b$ to any $1$-cycle $z$ yields another
$1$-cycle, $z+b$, but in enumerating gauge invariants, we should not
count $z$ and $z+b$ separately.  Thus $b$ acts very much like
the $1$-boundary considered in topology.  In fact, \eqref{trivialcycle}
is exactly the definition of the boundary of the $2$-chain
$\bold k[g|h]$:
\begin{equation}
\partial\bold k[g|h] ~\equiv~ h\bold k[g] - \bold k[gh] + \bold k[h]
\rlap{\quad.}
\label{delkgh}
\end{equation}

Thus, to determine all non-trivial gauge invariants of a group acting
on a lattice, we calculate the group of $1$-cycles and divide by
the group of boundaries of $2$-chains; the quotient group is
called $H_1(G,L)$ ($G$ is the point group, $L$ the lattice).
As we have noted \cite{rabsonfisher02}, this
is less elegant than Mermin's method but, obviating the need
for a clever choice of gauge, has proven useful in establishing
theorems and generalizing results to modulated crystals.  Moreover,
the systematic way in which gauge invariants fall out of this
formulation has enabled us to find a new type, not previously
noted \cite{fisherrabsoninprep}.

A comparison to the definition of the boundary of the 2-simplex
in the second line of \eqref{partial1} reveals nearly the
same formal structure under the replacement $g\rightarrow(x_0,x_1)$,
$h\rightarrow(x_1,x_2)$, $gh\rightarrow(x_0,x_2)$, the difference
being that we considered topological chains over integers,
while the coefficients in crystallography lie instead
in the lattice, on which the group acts non-trivially: hence
the $h\bold k$ in the first term on the right instead of just $\bold k$.

There is a way of visualizing $1$-chains in group homology that emphasizes the
analogy with $1$-cycles in topology:
picture the $1$-chain~$\bold k[g]$ as a vector going from~$\bold k$
to~$g \bold k$.  If $g$~is a rotation, it is suggestive to draw the vector
along a circular arc, as in Figure~5, in which $r$~denotes a
$90^\circ$~rotation.  This visualization is consistent with the
definition~\eqref{partialkg} of the boundary of a $1$-chain.  The
formula~\eqref{delkgh} for a $1$-boundary can be written in the form
$ \bold k[gh] \doteq \bold k[h] + (h \bold k)[g] $,
where $\doteq$ denotes homology as in \S\ref{topo}.  This is also consistent
with the vector picture:  for example, Figure~5 illustrates the fact that
$ \bold a[r^3] \doteq \bold a[r] + (r \bold a)[r^2] $.
One shortcoming of this method is that it does not illustrate linearity
relations, such as
$ (\bold a + \bold b)[g] = \bold a[g] + \bold b[g] $
and
$ (-\bold a)[g] = -(\bold a[g]) $.

\def\captionfive{%
It may help to elucidate chains to think of them geometrically;
a chain $\bold k[g]$ can be pictured as a circular arc pointing
from $\bold k$ to $g\bold k$.
The boundary $\partial\bold a[r^3] = \partial\bold a[r^{-1}] \doteq
\bold a[r] + (r\bold a)[r^2]$, as suggested by the three arcs.
}
\begin{figure}[tp]
\centerline{\epsfxsize0.5\hsize\epsfbox{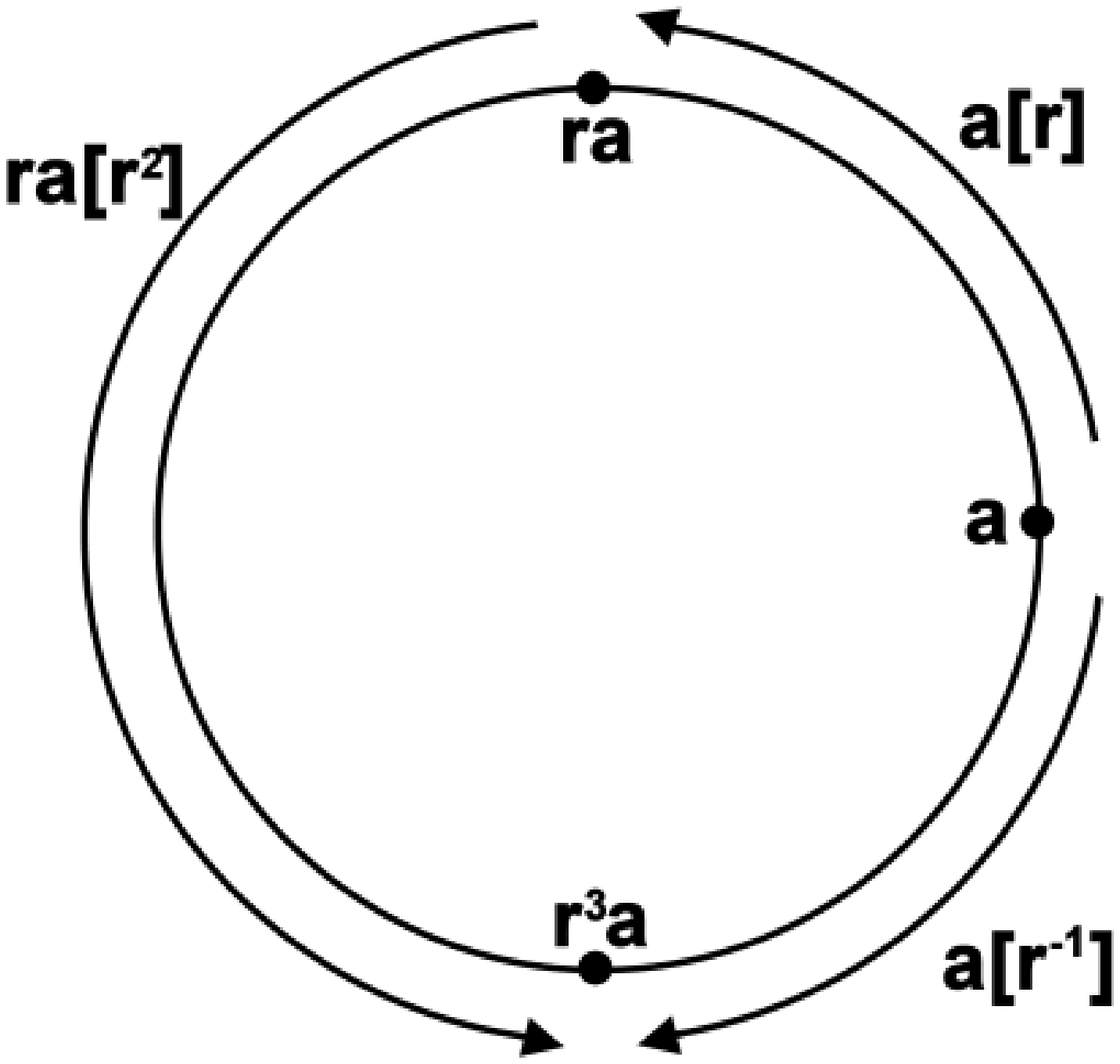}}
\caption[Figure 5]{\captionfive}
\end{figure}

\section{An Example of Calculating Space (Plane) Groups}
\label{examplesect}

To illustrate the homological method, we will calculate the homology
group $H_1(G,L)$ where the primitive rectangular lattice $L$ is
generated by a vector $a$ in the $\hat{\bold x}$ direction and
a vector $b$ of different length in the $\hat{\bold y}$ direction and
where the point group
$G=\{e,r,m,rm\}$ is generated
by a $180^\circ$ rotation $r$ and the mirror $m$ that leaves $a$ invariant
($e$ is the identity).
According to the International Tables for Crystallography
\cite{International87}, the possible
plane groups are $p2mm$, $p2gg$, and $p2mg$.  We shall verify this result in
the next subsection; we concentrate first on calculating the possible
invariants.

Since our goal is to illustrate a general method as simply as
possible, we avoid the shortcuts we used
in the more complicated example of section 7
of \cite{rabsonfisher02}.
In section \ref{mathemagica}, we show
how the entire calculation can be automated in the symbolic-algebra
program, {\itshape Mathematica} \cite{wolfram}, for arbitrary examples.

\subsection{Invariants of the Point Group {\itshape 2mm} on the Primitive
Rectangular Lattice}
\label{p2mm}
There are four elements of $G$, and   the lattice is generated by
two vectors.    To find all the cycles, we therefore take boundaries of
the eight $1$-chains~$\bold k[g]$ where $\bold k$ is a lattice generator.
(Any other $1$-chain is an integral linear combination of these.)

\begin{equation}
\renewcommand{\arraystretch}{1.4}
\arraycolsep.5\arraycolsep
\begin{array}{r c r @{\hspace{2em}} r c r }
\partial a[e] & = & 0 & \partial b[e] & = & 0 \\
\partial a[m] & = & 0 & \partial b[m] & = & -2b \\
\partial a[rm] & = & -2a & \partial b[rm] & = & 0 \\
\partial a[r] & = & -2a & \partial b[r] & = & -2b
\end{array}
\label{boundaries1}
\end{equation}

\relax From these eight values, we read off the generators of the
additive group of cycles (the angle brackets mean ``generated by''):

\begin{equation}
Z_1(G,L) = \langle z_1,\,z_2,\,z_3,\,z_4,\,z_5,\,z_6 \rangle
\label{Z1}
\end{equation}
where
\begin{equation}
\renewcommand{\arraystretch}{1.4}
\arraycolsep.5\arraycolsep
\begin{array}{r c r @{\hspace{3em}} r c l }
z_1 &=& b[e] & z_4 &=& a[e] \\
z_2 &=& -b[r] + b[m] & z_5 &=& a[m] \\
z_3 &=& b[rm] & z_6 &=& a[rm]-a[r]\rlap{\smash{\quad.}}
\end{array}
\label{Z1z}
\end{equation}

Now, we write down the 32 generators
$ \bold k[g|h] $
($\bold k = a$,~$b$ and $g$,~$h$ in the point group)
of all $2$-chains, and calculate
their boundaries:

\begin{equation}
\renewcommand{\arraystretch}{1.4}
\arraycolsep.5\arraycolsep
\begin{array}{r c c c l}
b_1 &\equiv& \partial a[e|e] &=& \hphantom{-}a[e]\\
b_2 &\equiv& \partial b[e|e] &=& \hphantom{-}b[e]\\
b_3 &\equiv& \partial a[r|m] &=& \hphantom{-}a[r] + a[m] - a[rm]\\
b_4 &\equiv& \partial a[m|r] &=& \hphantom{-}a[r] - a[m] - a[rm]\\
b_5 &\equiv& \partial b[r|m] &=& -b[r] + b[m] -b[rm]\\
b_6 &\equiv& \partial b[m|r] &=& \hphantom{-}b[r] - b[m] - b[rm]
\rlap{\quad.}
\end{array}
\label{boundaries2}
\end{equation}
All of the remaining boundaries are linear combinations of these six boundaries.
Furthermore, it is evident that the six boundaries in \eqref{boundaries2}
are integrally linearly independent, so%
\footnote{In \S \ref{section:finite}, we prove that $Z_1$~and~$B_1$ always
have the same rank.}
\begin{equation}
B_1(G,L) = \langle b_1\,,b_2\,,b_3\,,b_4\,,b_5\,,b_6 \rangle\rlap{\quad.}
\label{B1}
\end{equation}

We expect and verify that every boundary in \eqref{B1}
can be written as an integral linear
combination of the cycles in \eqref{Z1z} (since all boundaries are cycles).
However, the cycles $z_1$ and $z_4$ are actually boundaries, so we
throw them out.  Furthermore, $z_5-z_6=b_3$, which is a boundary,
so we write $z_5\doteq z_6$ (equality up to boundaries); similarly,
$z_2\doteq z_3$.  This leaves only two cycles, $z_3$ and $z_5$, which
are obviously linearly independent and not boundaries.  Finally, we note that
$2z_3 = -(b_5+b_6)$ and
$2z_5 = b_3 - b_4$ are boundaries.  Thus there
are only four invariants:
\begin{equation}
H_1(G,L) = \{ 0, \, z_3, \, z_5, \, z_3 + z_5 \} \rlap{\quad.}
\label{H1GL}
\end{equation}

\subsection{Connection between invariants and space groups (cohomology)}
\label{connection}
Now we're ready to classify plane groups.  According to the
Rokhsar-Wright-Mermin specification, we need to find all
functions $\Phi$, linear modulo unity on the lattice, satisfying
\eqref{compat}.  Since $2z_3$~and~$2z_5$ are boundaries, we must have
$2\Phi(z_3)=2\Phi(z_5)=0$ (all modulo unity), so that
there are four
possible gauge-invariant values for the phase function, given in
Table 2.

\begin{table}
\begin{center}
\renewcommand{\arraystretch}{1.1}
\begin{tabular}{| c | c | c |}
\hline
possibility & $\Phi(z_3) = \Phi_{rm}(b)$ & $\Phi(z_5) = \Phi_m(a)$ \\
\hline
1 & 0 & 0 \\
\hline
2 & 1/2 & 0 \\
\hline
3 & 0 & 1/2 \\
\hline
4 & 1/2 & 1/2 \\
\hline
\end{tabular}
\end{center}
\caption[Table 2]{The four sets of gauge-invariant values
a phase function might take for the point group $2mm$ on the
primitive-rectangular lattice.  These correspond to the plane
groups $p2mm$, $p2mg$, $p2gm$ (which is just a different setting
of $p2mg$),
and $p2gg$.}
\end{table}

We now make two assertions, the proof of which is the main content of
reference \cite{rabsonfisher02}.  First, two phase functions 
not related by a gauge differ in their values on at least one
gauge invariant in $H_1(G,L)$.  Thus, there are no more than four
space groups in the example.  Second, for each possibility, a
phase function exists.

The proof in \cite{rabsonfisher02} rests on the duality of homology
and cohomology, in which $\Phi$ plays the same role with respect to
a chain $\sum\bold k[g]$ as the vector field $\mathbf{F}$ played with respect
to a contour in subsection
\ref{deRham}.
Think of
$\Phi$ as acting on $g$ to give $\Phi_g$; $\Phi_g$ acts linearly on
$\bold k\in L$ to give the number $\Phi_g(\bold k)$, modulo unity as always.
The
$1$-cochains
form a group under addition.  A $1$-cochain is a \textit{$1$-cocycle} if its
coboundary, defined by
\begin{equation}
(d\Phi)(g,h) = \Phi_g \circ h - \Phi_{gh} + \Phi_h\rlap{\quad,}
\label{dPhi}
\end{equation}
vanishes: this is simply the group-compatibility condition on $\Phi$,
so we are clearly interested in $1$-cocycles.  Note that where the
boundary operator $\partial$ decreased the number of group arguments,
the coboundary operator $d$ increases it.  A linear function $\chi$ on the
lattice is called a \textit{$0$-cochain}; its coboundary is given by
\begin{equation}
(d\chi)(g) = \chi \circ g - \chi\rlap{\quad.}
\label{dchi}
\end{equation}
Two different $1$-cochains $\Phi$ and $\Phi'$ that differ by a coboundary
of the form of \eqref{dchi} are related by a gauge function, according
to \eqref{gauge}, so we identify them.  The group of $1$-cocycles modulo
$1$-coboundaries, or first cohomology, is labeled $H^1(G,\hat L)$, where
$\hat L$ denotes the group of linear maps from $L$ to $\R/\Z$;
if $\chi$~is an element of~$\hat L$, then
$ 2 \pi i \chi(\bold k) $
is a choice of phase for each $\bold k$ among a generating set for
the lattice $L$.
Table 3 summarizes
homology and cohomology; note the similarity between the requirement
for homology and triviality for cohomology, and {\itshape vice-versa}.
\begin{table}
\begin{center}
\renewcommand{\arraystretch}{1.1}
\tabcolsep1.5\tabcolsep
\begin{tabular}{|c|c|c|}
\hline
& homology & cohomology \\
\hline
1-(co)chain & $\sum_i\bold k_i[g_i]$ & $\Phi$ \\
1-(co)cycle & gauge invariant & satisfies group compatibility \\
1-(co)boundary & trivial by group compatibility & trivial by gauge
transformation\\
\hline
\end{tabular}
\end{center}
\caption[Table 3]{Summary of the application of group homology and
cohomology to Fourier-space crystallography.  (For additional applications
and higher-order (co)homology, see \cite{rabsonfisher02,FisherRabson01}.)
Symbols: $\bold k_i$ are reciprocal-lattice vectors and $g_i$ point-group
elements.  $\Phi$ is a phase function.  Compare
Table 1.}
\end{table}

The task of classifying space
groups, then, comes down to calculating the cohomology group of
cocycles modulo coboundaries, and this group is dual to the homology
group of cycles modulo boundaries.  Thus, the four possibilities in
Table 2 above are exactly the four possible space groups for this example.

However, there are only three space groups for $2mm$ on the primitive
rectangular lattice: $p2mm$, $p2mg$, and $p2gg$.  Possibilities
2 and 3 in the table simply exchange the $\hat{\bold x}$ and $\hat{\bold y}$
directions.\footnote{%
Once cohomology has been calculated, it is
still necessary to consider rotations and (in quasiperiodic cases)
scale invariance of the lattice.
See \cite{Dr-M,lifshitz96} for a full accounting of {\itshape Bravais class},
{\itshape arithmetic crystal class}, and {\itshape space-group type}.
}

\subsection{Comparison to torus and lemniscate}
\label{lemniscate}
Table 2 shows that
the homology (or cohomology) group of the point group~$2mm$
acting on the primitive-rectangular lattice
is isomorphic to
$\Z_2\times\Z_2$, the same homology group we considered in two
examples in section \ref{topo}.\footnote{To the manifold with two holes and
the torus, each with coefficients in $\Z_2$, we may
add $\R P^2\times\R P^2$ with coefficients in $\Z$.
}
The topological examples contained
two holes in the sense of closed curves that did not bound:
Figure~1 without orientation has two literal holes, while the
doughnut has
the hole in the
center and the fried dough (which is not part of the surface and so
is no different from a hole).
In each case, exchanging the two holes reduces the number of
combinations from four to three.  The close analogies tell us that
we can consider crystallography, as well, in terms of topological spaces.
A point group and lattice admitting no non-symmorphic space groups is a
trivial space, in which every closed curve
can be collapsed to a point, while non-symmorphic space groups are
non-zero cohomology classes in spaces with
one or more holes.  We shall show in section \ref{mathemagica} that
in every case, the invariants correspond to some number of holes, each of which
can be lassoed by a cycle a finite number of times before it vanishes.
(More prosaically, the (co)homology group is isomorphic to the direct
product of some number of cyclic groups of varying order.)

\section{Generalizing the calculation}

\subsection{Automation of calculation in {\itshape Mathematica}}
\label{mathemagica}
The algorithm of section \ref{p2mm}, requiring no clever choices of gauge or
anything specific to the point group or lattice, is easily automated.  Our
{\itshape Mathematica\/} implementation, available at
the preprint archive, {\ttfamily cond-mat/0301601},
takes only
13 lines of substantive code.  A quick tour of the algorithm may
assist those trying to do the calculation by hand as well as users
of computer-algebra packages.

\begin{enumerate}
\item
\label{foo1}
The point-group generators and their actions (to the right)
on the lattice are specified by square matrices of dimension equal
to the rank of the lattice.  For example, the two-fold rotation $r$
and mirror $m$ in section \ref{p2mm} are the matrices
\begin{equation}
r = \left(\!\!
\begin{array}{rr}
-1 & 0\\
0 & -1
\end{array}
\right);
\quad
m = \left(
\begin{array}{rr}
1 & 0\\
0 & -1
\end{array}
\right)
\label{p2mm:rm}
\end{equation}
acting on the column vectors
$a=\bigl( {1 \atop 0} \bigr)$
and
$b=\bigl( {0 \atop 1} \bigr)$.

\item
\label{foo2}
The group~$C_1$ of all 1-chains is generated by the
combinations~$\bold k[g]$, where $\bold k$~is a lattice generator and
$g\in G$. The computer
stores all these combinations in the list~$\ttm c$.
Think of this list, say
$ {\ttm c} = (c_1, \ldots, c_r) $,
as a column vector.

\item
\label{Z1step}
\label{foo3}
Take the boundary \eqref{delc}
of each~$c_i$.  The {\ttfamily NullSpace[\,]} function in
{\itshape Mathematica\/}, applied to the list of boundaries, gives
a matrix $\ttm z$ that expresses generators~$z_i$ of the cycle group~$Z_1$ in
terms of~$\ttm c$:
\begin{equation}
	\begin{pmatrix}
		z_1 \\ \vdots \\ z_s
	\end{pmatrix}
	= {\ttm z}
	\begin{pmatrix}
		c_1 \\ \vdots \\ c_r
	\end{pmatrix} .
\end{equation}

\item
\label{foo4}
The group~$C_2$ of all 2-chains is generated by the
combinations~$\bold k[g|h]$, where $\bold k$~is a lattice generator and
$g$,~$h\in G$.
The boundaries~$b_i$ of these 2-chains \eqref{delkgh} generate the
group~$B_1$ of $1$-boundaries.  Express these generators in terms of~$\ttm c$:
\begin{equation}
	\begin{pmatrix}
		b_1 \\ \vdots \\ b_t
	\end{pmatrix}
	= \ttm{bz3}
	\begin{pmatrix}
		c_1 \\ \vdots \\ c_r
	\end{pmatrix} .
\end{equation}
\label{stepB1}

\item
\label{foo5}
Find a rational transformation~$\ttm{z2}$ that transforms a cycle from
the basis of $C_1$ into the basis of $Z_1$. Our package uses
{\ttfamily PseudoInverse[\,]}.
\label{stepB1Z1}

\item
\label{foo6}
\relax From steps \ref{stepB1} and \ref{stepB1Z1},
the product $\ttm{bz3}.\ttm{z2}$ expresses
the generators~$b_i$ of $B_1$
in the basis of $Z_1$. We call {\ttfamily LatticeReduce[\,]}
to find the smallest set of rows that will generate $B_1$ over
the integers, yielding the matrix~$\ttm b$.
Thus
\begin{equation}
	\begin{pmatrix}
		b_1 \\ \vdots \\ b_s
	\end{pmatrix}
	= {\ttm b}
	\begin{pmatrix}
		z_1 \\ \vdots \\ z_s
	\end{pmatrix}
\end{equation}
gives a minimal set of generators for~$B_1$.
We explain in \ref{section:finite} below that $B_1$~and~$Z_1$ have the same
rank, so that $\ttm b$~is a square matrix.
We can calculate the first homology group
$ H_1 = Z_1 / B_1 $
as follows.
The Smith normal form
\cite[Thm.\ 25.26]{Spindler94}, \cite{Jabon94} of
the matrix~$\ttm b$ is a diagonal matrix~$\ttm d$ such that there are
invertible integral transformations $\ttm p$ and $\ttm q$ with
\begin{equation}
\ttm d ~=~ \ttm p \, \ttm b \, \ttm q\rlap{\quad.}
\label{smith}
\end{equation}
Thus $\ttm q$~and~$\ttm p$ describe
new bases for $Z_1$ and $B_1$, say $\{z'_i\}$~and~$\{b'_i\}$.
In these
new bases,
$ b'_i = d_i \, z'_i $;
in other words, $d_i$ times cycle $z'_i$ is a boundary.
Thus the homology group~$H_1$ can be described as all linear combinations of
$z'_1$, ~$\ldots$,~$z'_s$,
where the coefficient of~$z'_i$ lies between $0$~and~$d_i-1$, and there are no
relations other than $d_i z'_i \doteq 0$.

\item
\label{foo7}
Our function {\ttfamily SpacegroupH1Full[\,]}
returns the matrices $\ttm c$,
$\ttm z$, $\ttm b$, $\ttm d$,
$\ttm p$, and $\ttm q$ above.
\end{enumerate}

\begin{table}
\begin{center} 
\def\circle#1{\rlap{\hbox to0.25em{}#1}\raise0.075em\hbox{$\bigcirc$}}
\def\foo#1{\circle{\ref{foo#1}}}
\catcode`\!0
\begin{verbatim}
            (* For g a list of matrices, return <g>, the group generated. *)
         groupgen[g_] := FixedPoint[Union[Flatten[Outer[Dot,#,g,1],1],#]&, g]
            (* Project a term of form (v1 y[g1]) onto elements of c1. *)
         c1proj[x_,c1_] := ( (x/.{y[g_]->If[g==#[[2]],1,0]}).#[[1]] )& /@ c1
    !foo1   latnum = Dimensions[x[[1]][[1]]][[1]]; (* x = list of generators *)
         latgen = IdentityMatrix[latnum];
         group = groupgen[x];
    !foo2   c = Distribute[{latgen,group}, List];
    !foo3   z = NullSpace[Transpose[(#[[2]].#[[1]] - #[[1]])& /@ c]];
    !foo4   c2 = Distribute[{latgen,group,group}, List];
         bz2 = Union[((#[[3]].#[[1]])y[#[[2]]] - #[[1]]y[#[[2]].#[[3]]] +
         	#[[1]]y[#[[3]]])& /@ c2];
         bz3 = Union[(c1proj[#,c])& /@ bz2];
    !foo5   z2 = Chop[Rationalize[PseudoInverse[N[z]]]];
    !foo6   b = LatticeReduce[Rationalize[Chop[bz3.z2]]];
         {d,{p,q}}=ExtendedSmithForm[b];
\end{verbatim}
\end{center}
\caption[Table 4]{The substantive lines of {\itshape Mathematica\/}
code described in section \ref{mathemagica}.  The circled numbers
refer to the steps listed in the text.  The full version of the
code is available at the preprint archive, {\ttfamily cond-mat/0301601}.
}
\end{table}

Returning again to the example of section \ref{p2mm}, we find
\begin{equation}
\ttm d = \hbox{diag}(1,1,1,1,2,2)\rlap{\quad.}
\end{equation}
The module $Z_1$ has dimension~six, as in \eqref{Z1z}.
The four unit entries indicate that the corresponding cycles
are boundaries,
while the two entries
of $2$ tell us that
two, $z'_5$ and $z'_6$,
are not, although $2z'_5$ and $2z'_6$ are.
We invert the Smith transformation to solve for $z'_5$ and $z'_6$:
\begin{equation}
\ttm p^{-1}~\hbox{diag}(0,0,0,0,1,1)~\ttm q^{-1}\rlap{\quad.}
\label{PQinverse}
\end{equation}
This
yields a matrix with two non-zero rows, $(0,-1,0,0,0,0)$ and $(0,0,0,0,0,1)$,
giving $z'_5=-z_2$ and $z'_6=z_6$.
The $\ttm z$ matrix converts the $Z_1$ basis into the basis of $C_1$, which
we decode with the help of the list $\ttm c$ to compare with the
results of section \ref{p2mm}:
\begin{equation}
\begin{aligned}
z'_5&=&-z_2~\doteq~z_3\\
z'_6&=&\hphantom{-}z_6~\doteq~z_5\rlap{\quad.}
\label{comparez}
\end{aligned}
\end{equation}

We admit this algorithm to be inelegant: like
direct-space formulations,
it requires the construction
of matrices of dimension equal to the rank of the
lattice.  Unlike the Rokhsar-Wright-Mermin method,
it cannot treat related point groups, {\itshape e.g.},
$2^jmm=\{4mm,\,8mm,\,(16)mm\,\dots\}$, all at once \cite{Rabson89} but
rather requires a separate calculation for each.  However,
we find it useful in those cases (such as 46-fold
symmetry and modulated lattices) where discovering a good
choice of gauge, requisite to Rokhsar-Wright-Mermin, seems too difficult.

\subsection{Finite number of Space Groups}
\label{section:finite}
When applying the Rokhsar-Wright-Mermin approach to quasicrystals, there are a
few surprises.  One is that, even in two dimensions, a
finitely-generated lattice that is not discrete may have an infinite point
group~\cite{Piun-survey}.
Assume, as we have been doing implicitly so far, that the point group is
finite.  Happily, this is enough to guarantee that there are only finitely
many space groups associated to the point group~$G$ and the lattice~$L$.
As explained in \S\ref{connection}, this amounts to saying that the homology
group
$ H_1(G, L) $
is finite.  In any example, this can be checked by verifying that the matrices
$\ttm b$~and~$\ttm d$
of~\S\ref{mathemagica} have the same rank as the cycle group~$Z_1$.

In order to prove generally that
$ H_1(G, L) $
is finite, it suffices to show that $Nz$~is a boundary for any $1$-cycle~$z$,
where $N$~is the order of the point group~$G$.  Grant this for the moment.
There are then
finitely many $1$-cycles, say
$z_1$,~$z_2$, \dots,~$z_s$,
(where $M$~is the rank of the cycle group~$Z_1$)
that generate all the others:  any $1$-cycle~$z$ can be written as a linear
combination of the generators~$z_i$.  In order to represent all homology
classes, it suffices to take coefficients between $1$~and~$N$, since
$Nz_i$~is a boundary.
Thus there are only finitely many homology classes.
We can also explain this in terms of matrices:  if $N z_i$~is a boundary for
each generator~$z_i$, then
$ {\ttm b'} \, {\ttm b} = N I_M $
for some integer matrix~$\ttm b'$, where $I_s$ denotes the $s\times s$
identity matrix.
This implies that $\ttm b$~has rank~$s$.

It remains to explain why $N$~times a cycle is a boundary.
Let
$ z = \sum_h \bold k_h[h] $
be a cycle, where $h$~runs through the point group.  According to
\eqref{delc}, the cycle condition
$ \del z = 0 $
can be written in the form
\begin{equation}
	\label{eqn:altcycle}
	\sum_h h \bold k_h = \sum_h \bold k_h .
\end{equation}
Construct a $2$-chain~$c$ such that
$ \del c = N z $
using the fundamental technique of averaging over the group:  let
\begin{equation}
	c = \sum_{g,h} \bold k_h[g|h] .
\end{equation}
According to~\eqref{delkgh},
\begin{equation}
	\label{eqn:average}
	\del c = \sum_{g,h} (h \bold k_h)[g]
	- \sum_{g,h} \bold k_h[gh]
	+ \sum_{g,h} \bold k_h[h] .
\end{equation}
The first two terms in~\eqref{eqn:average} cancel:  to see this, fix~$g$ and
apply the cycle condition~\eqref{eqn:altcycle} in the first term; and fix~$h$
and replace~$g$ with~$gh^{-1}$ in the second term.  The summand in the third
term in~\eqref{eqn:average} is independent of~$g$, so the third term is
simply~$Nz$.

\subsection{From Real Space to Fourier Space}
So far, we have described the Fourier-space approach to crystallography in the
language of (co)homology, explained how this language lends itself to explicit
calculations, and pointed out the analogy with topological (co)homology.  The
real advantage of adopting this framework comes from applying the theorems
developed over the years for homological algebra in general and group
cohomology in particular.
The duality between phase functions and crystallographic invariants
(\S\ref{section:duality}) is a special case of one such theorem, the duality
between homology and cohomology.  The calculation of invariants
(\S\S~\ref{p2mm} and~\ref{mathemagica}) is one example of a standard
technique.  The finiteness result (\S\ref{section:finite}) is another.

In this subsection, we summarize the connection between the real-space and
Fourier-space approaches to crystallography, using the common language of
group cohomology.  Each step in this comparison involves a standard result,
explained in textbooks such as~\cite{Brown}, but we will not try to reproduce
these explanations.  At the end, we illustrate these generalities by returning
to the example of the point group~$2mm$.

In real-space crystallography, one considers a crystal and its \textit{space
group}~$\G$.  For a periodic, $d$-dimensional crystal, the space group
consists of all isometries of~$\R^d$ that preserve the crystal.  If we
are interested in a quasiperiodic crystal, we construct a periodic
crystal in a higher-dimensional
space, say~$\R^D$, that projects to the quasiperiodic one;
for this discussion, $\G$ will be the
space group of the higher-dimensional, periodic crystal.\footnote{%
Different choices of the periodic $D$-dimensional crystal may lead
to different $D$-dimensional space groups \cite{RWM88b}.
}
(In the periodic case, take $D=d$.)
Let $\T$ denote the lattice of pure real-space translations in $\G$.

In group-theoretic terms, the point group
is the quotient $ G = \G / \T $.
It can be thought of as the group of ``macroscopic
symmetries'':  think of all translations as ``microscopic,'' and so identify
any two elements of the space group if they differ by a translation.
Both $G$~and~$\T$ are fairly easy to describe:
$G$~is a finite subgroup of the orthogonal group,%
\footnote{The real-space approach cannot be used to describe quasicrystals with
infinite point groups, but it is not clear that these are physically
interesting.}
and $\T$~consists of all integral linear combinations of $D$~generating
vectors.
These two groups are also
fairly easy to determine experimentally.  It remains
to describe the space group~$\G$ in terms of these two simpler groups.
One possibility is that the point group is contained in~$\G$, and the space
group is generated by $G$ and~$\T$.  In this case, the space group is called
\textit{symmorphic} in crystallographic terminology, or a \textit{semidirect
product} in the language of group theory.

Not all space groups are symmorphic.  A standard result of group theory states
that the space groups corresponding to $G$~and~$\T$ are classified by the
cohomology group%
\footnote{This result applies because $\T$ is a normal, Abelian subgroup of
$\G$.}
$ H^2(G, \T) $.
Using the Long Exact Sequence of cohomology
(\cf~the last Remark in \S5 of~\cite{Hiller}),
this is isomorphic to
$ H^1(G, \R^D / \T) $.

In the periodic case, the Fourier lattice~$L$ is the dual of the translation
group~$\T$.  That is,
the Fourier lattice consists of those vectors $\bold k$ such that
$\bold k\cdot\bold t$ is integral for all $\bold t$ in the direct
lattice $\T$.
(Those who worry about where the factor of~$2 \pi$ went should refer to
\eqref{densities}~and~\eqref{phase}.)
Dr\"ager and Mermin~\cite{Dr-M} realized that this can be turned around in the
quasiperiodic case:  the translation group~$\T$ and the super-space~$\R^D$ can
be described, without explicit coordinates, as duals of the Fourier lattice.
Algebraically, this leads to the isomorphism
$ \R^D / \T = \hat L $,
where $\hat L$~denotes the linear maps from~$L$ to~$\R/\Z$ as
in~\S\ref{connection}.
Thus space groups are classified by the cohomology group
$ H^1(G, \hat L) $,
which is exactly the set of phase functions (identifying those that differ by
a gauge transformation) described in~\S\S\ref{section:FSC}
and~\ref{connection}.

We now return to the point group~$2mm$ as in \S\ref{p2mm}.
Let $\T$ denote the lattice in~$\R^2$ dual to the Fourier lattice~$L$
generated by $a$~and~$b$.
We will use the terminology of cocycles and coboundaries introduced
in~\S\ref{connection}, with $\Phi$~replaced by~$\bold t$.
We will also use the notation~$\spg{g}{\bold t}$ for elements of the space
group, where $g$~is in the point group and $\bold t$~is a translation (not
necessarily in~$\T$):  this acts on the vector~$\bold x$ by the formula
$ \spg{g}{\bold t} \bold x = g \bold x + \bold t $.
It follows that
$ \spg{g_1}{\bold t_1} \cdot \spg{g_2}{\bold t_2}
	= \spg{g_1 g_2}{\bold t_1 + g_1(\bold t_2)}
$.

For each element~$g$ of
$ G = \{e, r, m, rm\} $,
choose a translation~$\bold t_g$ such that
$ \spg{g}{\bold t_g} $
is in the space group.
By choosing the origin appropriately, we can arrange it so that
$\bold t_m$~lies along the $\ivec$-axis
and
$\bold t_{rm}$~lies along the $\jvec$-axis.
For each of the glide reflections
$ \spg{m}{\bold t_m} $~and~$ \spg{rm}{\bold t_{rm}} $,
there are two possibilities:  the translational part can lie in~$\T$, or it
can be half of a translation in~$\T$.
Since $m$~and~$rm$ generate the point group, the choice of
$\bold t_m$~and~$\bold t_{rm}$
determines the space group.
We shall see that the different possible choices lead to the four possibilities
in Table~2.
We shall postpone the verification that each of the four choices actually 
leads to a space group with translation subgroup~$\T$ and point group~$G$.

The definition of multiplication in the space group shows that the combination
(\cf~\eqref{dPhi})
\begin{equation}
	\label{eqn:dt}
	d \bold t(g,h) \equiv \bold t_g - \bold t_{gh} + g(\bold t_h)
\end{equation}
is the translational part of
$ \spg{g}{\bold t_g} \spg{h}{\bold t_h} \spg{gh}{\bold t_{gh}}^{-1} $
and so lies in~$\T$.  In fact, $d \bold t$~is a $2$-cocycle with values in~$\T$.%
\footnote{If we allow cochains with values in~$\R^2$, then $\bold t_g$~is a
$1$-cochain and
$d \bold t$~is its boundary.  It is therefore automatically a cycle.  However,
$d \bold t$~is not necessarily the boundary of a $1$-cochain with values
in~$\T$.}
The choice of origin, mentioned above, does not change the value
of~$d \bold t$.  Choosing a different translation~$\bold t_g$ for each~$g$
would change~$d \bold t$ by a $1$-coboundary, so it makes sense to identify 
cocycles that differ by a coboundary.  Thus $d \bold t$~determines a
class in
$ H^2(G, \T) $,
and this cohomology class corresponds to the given space group.

Another point of view is that one should not choose a particular~$\bold t_g$
but should consider
all possible choices at once.  In other words, instead of the
vector~$\bold t_g$, consider its coset
$ \bar{\bold t}_g = \bold t_g + \T $,
as an element of the quotient group~$\R^2 / \T$.
\relax From this point of view, $\bar{\bold t}$~is a $1$-cochain with values
in~$\R^2/\T$.  The fact that $d \bold t$~takes values in~$\T$ means that
$\bar{\bold t}$~is a cocycle.  A different choice of origin changes this
cocycle by a coboundary.  Thus the space group is described by the class
of~$\bar{\bold t}$ in
$ H^1(G, \R^2 / \T) $.
The relation between
$ \bar{\bold t} $~and~$ d \bold t $
illustrates the general isomorphism between
$ H^1(G, \R^D / \T) $
and
$ H^2(G, \T) $.

The phase function that describes the space group is simply
$ \Phi_g(\bold k) = \bold k \cdot \bold t_g $,
where the dot product defines the duality pairing between Fourier and real
space.  Note that changing~$\bold t_g$ by a translation in~$\T$ changes the
phase by an integer, which we always ignore.  This illustrates the general
isomorphism between
$ H^1(G, \R^D / \T) $
and
$ H^1(G, \hat L) $.
As promised, the four possible choices of
$ \bold t_m $~and~$ \bold t_{rm} $
indeed correspond to the four possibilities in Table~2.
\relax From either the
Fourier-space point of view or the real-space one, two of the four space
groups are equivalent under exchanging and rescaling the axes.

We already know, from~\S\ref{connection}, that
$ H^1(G, \hat L) $
contains exactly four elements.  Therefore, the isomorphism between
$ H^2(G, \T) $
(which classifies space groups from the direct-space point of view) and
$ H^1(G, \hat L) $
(the group of phase functions up to gauge equivalence)
gives a round-about verification of the point we omitted above:
all four ways of choosing the glide reflections
$ \bold t_m $~and~$ \bold t_{rm} $
actually lead to space groups with translation
subgroup~$\T$ and point group~$G$.

\subsection{Is there a topological space for every crystal class?}

We have offered an analogy between the classification of space
groups\footnote{That is, just that part of the classification that concerns
families of gauge-equivalent phase functions \cite{Dr-M}.}
by group cohomology and some topological problems on manifolds.
One can ask how deep the connection really goes.  We might wish to know
whether, for a point group $G$ with its action on a
lattice $L$, we could always find a topological manifold $M$ and
coefficient group $C$ such that
\begin{equation}
H_n(G,L) ~{ \buildrel ? \over \cong }~ H_n(M,C)
\label{deepnine}
\end{equation}
for {\itshape all\/} integers $n\ge0$.
We do not know, in general, whether such a manifold
exists.  However, there does exist a general construction for a topological
space so that \eqref{deepnine} works for $n=1$.  Since $H_1(G,L)$ is
isomorphic to a finite direct product of cyclic groups
$ \Z_N = H_1(L_N, \Z) $,
where $L_N$~denotes the \textit{lens space}
\cite[\S24.4]{Dubrovin84}, the corresponding
direct product of lens spaces has the desired first homology group
\cite[Prop.~0.8]{Brown}.

The topological view provides a picture of
what one is doing in the Rokhsar-Wright-Mermin formulation.  There,
the task of finding all gauge invariants is complicated first by
the possibility that these gauge invariants may be linear combinations
of phase functions, as in \eqref{2KM}, and second by the need to find
a clever choice of gauge to make all values that are not invariant,
or derivable from them, vanish.  Viewed cohomologically, the second
point is no concern, since the gauge-invariant values
completely determine the phase function \cite{rabsonfisher02}. The
first complication is also no concern in the homological approach;
the fact that \eqref{2KM} is a linear combination of terms simply
reflects an inopportune choice of basis vectors for the cycle group~$Z_1$.
A more convenient basis would include an invariant like
$\bold k_h[g]-\bold k_g[h]$,
and each basis element can be thought of as corresponding to a
hole in a topological space.

\thanks{We wish to thank
Edwin Clark, Mohamed Elhamdadi, and Veit Elser for useful conversations.
This work was supported by the National Science Foundation
through grants DMS-0204823 and DMS-0204845.  DAR is a Cottrell
Scholar of Research Corporation.}

\newbox\integersmithnormalform\setbox\integersmithnormalform%
\hbox{\footnotesize\ttfamily IntegerSmithNormalForm}

\bibliographystyle{amsplain}
\bibliography{Quasicrystal}

\vfil\eject

\centerline{\bfseries Figure Captions}
\bigskip
\bigskip

{
\parindent0pt
\parskip1\baselineskip

{\bfseries Figure 1. }\captionone

{\bfseries Figure 2. }\captiontwo

{\bfseries Figure 3. }\captionthree

{\bfseries Figure 4. }\captionfour

{\bfseries Figure 5. }\captionfive

}
\vfil\eject

\centerline{\epsfxsize\hsize\epsfbox{Figure1.eps}}
\vfil
\centerline{Figure 1.}
\vfil\eject
\centerline{\epsfxsize\hsize\epsfbox{Figure2a.eps}}
\vfil
\centerline{Figure 2a.}
\vfil\eject
\centerline{\epsfxsize\hsize\epsfbox{Figure2b.eps}}
\vfil
\centerline{Figure 2b.}
\vfil\eject
\centerline{\epsfxsize\hsize\epsfbox{Figure3.eps}}
\vfil
\centerline{Figure 3.}
\vfil\eject
\centerline{\epsfxsize\hsize\epsfbox{Figure4.eps}}
\vfil
\centerline{Figure 4.}
\vfil\eject
\centerline{\epsfxsize\hsize\epsfbox{Figure5.eps}}
\vfil
\centerline{Figure 5.}
\vfil\eject

\end{document}